\documentclass[twocolumn]{aastex62}
\usepackage{amsmath}
\usepackage{amssymb}

\hypersetup{linkcolor=red,citecolor=blue,filecolor=cyan,urlcolor=magenta}

\received{}
\revised{}
\accepted{}
\submitjournal{\apjl}

\newcommand{\g}{\text{g}}
\newcommand{\cm}{\text{cm}}
\newcommand{\km}{\text{km}}
\newcommand{\G}{\text{G}}
\newcommand{\Hz}{\text{Hz}}

\newcommand{\se}{\text{s}}
\newcommand{\yr}{\text{yr}}

\newcommand{\ms}{\text{ms}}
\newcommand{\fme}{\text{fm}}
\newcommand{\MeV}{\text{MeV}}
\newcommand{\rad}{\text{rad}}
\newcommand{\dyn}{\text{dyn}}

\newcommand{\CB}{\mathcal{B}}
\newcommand{\CR}{\mathcal{R}}

\newcommand{\col}{\textcolor{black}}

\shorttitle{Glitch rises as a test for rapid superfluid coupling in neutron stars}
\shortauthors{Graber et al.}

\begin{document}

\title{Glitch rises as a test for rapid superfluid coupling in neutron stars}

\correspondingauthor{Vanessa Graber}
\email{vanessa.graber@mcgill.ca}

\author[0000-0002-6558-1681]{Vanessa Graber}
\affil{Department of Physics and McGill Space Institute, McGill University, 3550 rue University, Montreal, QC H3A 2T8, Canada}

\author[0000-0002-6335-0169]{Andrew Cumming}
\affil{Department of Physics and McGill Space Institute, McGill University, 3550 rue University, Montreal, QC H3A 2T8, Canada}

\author{Nils Andersson}
\affil{Mathematical Sciences and STAG Research Centre, University of Southampton, Southampton SO17 1BJ, United Kingdom}

\begin{abstract}

Pulsar glitches provide a unique way to study neutron star microphysics because short post-glitch dynamics are directly linked to strong frictional processes on small scales. To illustrate this connection between macroscopic observables and microphysics, we review calculations of vortex interactions focusing on Kelvin wave excitations and determine the corresponding mutual friction strength for realistic microscopic parameters in the inner crust. These density-dependent crustal coupling profiles are combined with a simplified treatment of the core coupling and implemented in a three-component neutron star model to construct a predictive framework for glitch rises. As a result of the density-dependent dynamics, we find the superfluid to transfer angular momentum to different parts of the crust and the core on different timescales. This can cause the spin frequency change to become non-monotonic in time, allowing for a maximum value much larger than the measured glitch size, as well as a delay in the recovery. The exact shape of the calculated glitch rise is strongly dependent on the relative strength between the crust and core mutual friction, providing the means to probe not only the crustal superfluid but also the deeper neutron star interior. To demonstrate the potential of this approach, we compare our predictive model with the first pulse-to-pulse observations recorded during the December 2016 glitch of the Vela pulsar. Our analysis suggests \col{that the glitch rise behavior is relatively insensitive to the crustal mutual friction strength as long as $\CB \gtrsim 10^{-3}$, while being strongly dependent on the core coupling strength, which we find to be in the range $3 \times 10^{-5} \lesssim \CB_{\rm core} \lesssim 10^{-4}$.}
\end{abstract}



\section{Introduction}
\label{sec:intro}

Neutron stars provide the unique opportunity to study matter under extreme conditions. Learning about their unknown nuclear equation of state (EoS) relies on understanding the connection between the macroscopic observables and microphysics. One possibility is to probe the interior physics with \textit{glitches}. These sudden spin-ups interrupt the regular pulsar spin-down \citep{Espinoza2011} and are typically associated with the transfer of angular momentum from a crustal superfluid, decoupled from the lattice (and everything tightly coupled to it) due to vortex pinning \citep{Anderson1975}. Upon reaching a critical lag, the glitch is triggered and a large number of vortices simultaneously unpin. The frictional forces acting on free vortices on small scales and mechanisms causing their gradual repinning subsequently govern the macroscopic post-glitch response \citep{Pines1980}. The latter are typically associated with an exponential recovery and modeled within \textit{vortex-creep theory} \citep{Alpar1984b, Alpar1993, Akbal2017}, whereas the former dominate the behavior at early times. In this paper, we focus on the glitch rise.

Observations of the Vela pulsar suggest that crust coupling is very efficient: initial constraints for the spin-up timescale \citep{Dodson2002, Dodson2007} have been recently improved showing that the crust accelerates within $\sim 5 \, \se$ \citep{Palfreyman2018} after the glitch is initiated. Within hydrodynamical models, this rapid recoupling is captured via a dimensionless \textit{mutual friction} coefficient $\CB$ (directly related to the vortex dynamics) as the timescale to recouple the bulk superfluid is $\propto 1/ 2 \Omega_{\rm sf} \CB$ \citep{Alpar1988, Andersson2006b}. Provided that neutron stars are continuously monitored, spin-ups can be `caught in the act', allowing access to the early transient dynamics and the corresponding mutual friction coefficients, which in turn are controlled by the underlying small-scale processes. The most promising candidate to study this connection between macro- and microphysics is the Vela pulsar and dedicated observation campaigns have been performed, for example, at the Mount Pleasant Radio Observatory, Tasmania and the Hartebeesthoek Radio Astronomy Observatory, South Africa. Using the former, \citet{Palfreyman2018} have recently reported the first single-pulse observations of a sudden spin-up, providing the most detailed information of the glitch rise to date.

As a first step towards making realistic predictions for such an observation and constraining neutron star microphysics, we review two existing calculations \citep{Epstein1992, Jones1992} (as well as highlighting inconsistencies between them) analyzing the mechanism held responsible for rapidly recoupling the crustal superfluid -- excitation of \textit{Kelvin waves} along superfluid vortices. Instead of following previous work using constant mutual friction coefficients (e.g.~\citealt{Haskell2012}), we subsequently determine the Kelvin wave coupling strength for a realistic crust model, discussing uncertainties in the microscopic parameters. These new density-dependent couplings are further combined with a simplified treatment of the core coupling and implemented in a three-component neutron star model to make a prediction of the initial glitch response of a Vela-like pulsar. This is followed by a comparison between our predicted glitch rise and single-pulse observations of the December 2016 Vela pulsar glitch \citep{Palfreyman2018}.


\section{Rapid superfluid recoupling from Kelvin wave excitation}
\label{sec:coupling}

Following the large-scale unpinning initiating a pulsar glitch, \col{vortices move with a local velocity $\Delta v$ relative to the crustal lattice. Provided that $\Delta v$ is sufficiently large (see below for more details), the} excitation of circularly-polarized Kelvin waves dominates the dissipation. On small scales, these dynamics are fully characterized by a dimensionless drag parameter $\CR$, because an individual vortex feels a resistive force per unit length,
\begin{equation}
	 f_{\rm res} = \rho_{\rm s} \kappa  \CR \Delta v,
		\label{eqn-resforce}
\end{equation}
where $\rho_{\rm s} \equiv m_{\rm u} n_{\rm s}$ is the mass density of the free crustal superfluid, $m_{\rm u}$ the atomic mass unit, $n_{\rm s}$ the superfluid number density and $\kappa \approx 2.0 \times 10^{-3} \, \cm^2 \, \se^{-1}$ the quantum of circulation. Assuming that a large number of vortices moves freely and experiences $f_{\rm res}$, the microscopic drag is related to the large-scale hydrodynamic mutual friction coefficient by \citep{Glampedakis2011a}
\begin{equation}
	\CB \equiv \frac{\CR}{1 + \CR^2}.
		\label{eqn-defCB}
\end{equation}

To obtain $\CB$, the drag coefficient $\CR$ has to be known. Kelvin wave dynamics have been addressed by \citet{Epstein1992} and \citet{Jones1992}, albeit arriving at different results for the corresponding dissipation. In order to provide context for these papers and discuss the origin of the discrepancy, we use a simplified version of \citeauthor{Epstein1992}'s argument to derive the expected scalings for $\CR$. The equation of motion for forced vortex oscillations reads
\begin{equation}
	\rho_{\rm s} \kappa \hat{z} \times \frac{\partial \vec{\epsilon}} {\partial t} + T \, \frac{\partial^2\vec{\epsilon}}{\partial z^2} = \vec{f},
\end{equation}
where $\vec{\epsilon}$ is the displacement of a vortex aligned with the $z$-direction, $T$ the vortex tension and $\vec{f}$ the driving force per unit length. In the absence of forces, a plane wave ansatz shows that the vortex supports Kelvin waves with characteristic frequency \citep{Thomson1880, Jones1990}
\begin{equation}
	\omega_k = \frac{T k^2}{\rho_{\rm s} \kappa} = \frac{\hbar k^2}{2 \mu(k)}.
		\label{eqn:dispersionkelvon}
\end{equation}
Here, $k$ is the wave number along a vortex, $\hbar$ the reduced Planck constant and $\mu(k)$ an effective mass that varies slowly with $k$. This dispersion relation provides the tension associated with a specific mode, $T = \rho_{\rm s} \kappa\hbar/2\mu$.

Consider a point interaction with a lattice nucleus in which a force $f \sim \delta(z) E_{\rm p}/\ell$ is exerted on the vortex over a time $\tau \sim \ell/\Delta v$. $E_{\rm p}$ and $\ell$ are the pinning energy and typical interaction scale. This will excite Kelvin waves of characteristic frequencies $\omega \lesssim \tau^{-1}$ and wave numbers $k\lesssim k_* \equiv (2\mu/\hbar \tau)^{1/2}$, related to the Fourier-transformed amplitude $\tilde{\epsilon} (k \lesssim k_*) \sim E_{\rm p} \tau/ \rho_{\rm s} \kappa \ell$ (see also \citealt{Link2003}). The energy associated with the perturbations is
\begin{equation}
	\Delta E \sim \int T k^2 \tilde{\epsilon}(k)^2 \, {\rm d} k \sim \frac{\hbar k_*^3}{\rho_{\rm s}\kappa \mu} \left(\frac{E_{\rm p}} {\Delta v}\right)^{\! 2}.
\end{equation}
Since we are concerned with the scalings, numerical prefactors have been dropped. They are reintroduced below.

According to \citet{Epstein1992}, the power transferred into Kelvin waves per unit length is
\begin{equation}
	p \sim n_{\rm l} \Delta v \int \Delta E(b) db \sim \frac{\hbar k_*^3}{\rho_{\rm s} \kappa \mu} \left(\frac{E_{\rm p}}{\Delta v}\right)^{\! 2} n_{\rm l} \ell \Delta v,
\end{equation}
where we have $n_{\rm l}$ nuclei per unit volume and the integral over impact parameters $b$ is cut off at the scale $\ell$. A vortex hence experiences the resistive force $f_{\rm res} = p / \Delta v$ per unit length. With Equation~\eqref{eqn-resforce} and $k_*$, we obtain
\begin{equation}
  \CR \sim \left(\frac{\mu} {\hbar}\right)^{\! 1/2} \left(\frac{E_{\rm p}}{\rho_{\rm s} \kappa}\right)^{\! 2} \frac{1}{\Delta v^{3/2}} \frac{n_{\rm l}}{\ell^{1/2}}.
  \label{eqn-scalingR}
\end{equation}

\begin{deluxetable*}{c|ccccc|ccccc|cccccc}[t!]
\tablecaption{Composition for five crustal domains and corresponding vortex-nucleus interaction parameters. Baryon density $n_{\rm b}$, proton number $Z$, total neutron number $N$ within a Wigner-Seitz sphere, proton-to-neutron ratio $\tilde{x}$ inside a nucleus and free neutron density $n_{\rm s}$ are taken from \citet{Negele1973}. We calculate the total mass density $\rho \simeq m_{\rm u} n_{\rm b}$, number of baryons inside a nucleus
$A \simeq Z(1+1/\tilde{x})$, Wigner-Seitz radius $R_{\rm WS} \simeq [3(N+Z)/(4\pi n_{\rm b})]^{1/3}$, lattice nucleus density $n_{\rm l} \simeq 3/(4\pi R_{\rm WS}^3)$ and lattice constant $a \simeq (2/n_{\rm l})^{1/3}$. Estimates for the nuclear radius $R_{\rm N}$ and short-range
(long-range) contribution $E_{\rm s}$ ($E_{\rm l}$) to the pinning interaction are from \citet{Epstein1992}.\tablenotemark{a,b} Neutron gap $\Delta$ and coherence length $\xi$, related as $\xi = \hbar^2 k_{\rm Fs}/(\pi m_{\rm u}\Delta)$ ($k_{\rm Fs}$ is the free neutron Fermi wave number), and microscopic pinning energies $E_{\rm p}$ (corresponding to $\beta = 3$)
are taken from \citet{Donati2006}.\tablenotemark{c}
\label{tab-01}}
\tablecolumns{17}
\tablehead{
\colhead{} & \colhead{$n_{\rm b}$} & \colhead{$Z$} & \colhead{$N$} & \colhead{$\tilde{x}$} & \colhead{$n_{\rm s}$} & \colhead{$\rho$} & \colhead{$A$} &
\colhead{$R_{\rm WS}$} & \colhead{$n_{\rm l}$} & \colhead{$a$} & \colhead{$R_{\rm N}$} & \colhead{$E_{\rm s}$} & \colhead{$E_{\rm l}$} & \colhead{$\Delta$} &
 \colhead{$\xi$} & \colhead{$E_{\rm p}$} \\
\colhead{} & \colhead{[$10^{-4} \times$} & \colhead{ } & \colhead{ } & \colhead{ } & \colhead{[$10^{-4} \times$} & \colhead{[$10^{12}\times$} &
\colhead{} & \colhead{} & \colhead{[$10^{-6} \times$} & \colhead{} & \colhead{} & \colhead{} & \colhead{}
& \colhead{} & \colhead{} & \colhead{} \\
\colhead{} & \colhead{$\fme^{-3}$]} & \colhead{ } & \colhead{ } & \colhead{ } & \colhead{$\fme^{-3}$]} & \colhead{$\g\,\cm^{-3}$]} &
\colhead{} & \colhead{[$\fme$]} & \colhead{$\fme^{-3}$]} & \colhead{[$\fme$]} & \colhead{[$\fme$]} & \colhead{[$\MeV$]} &  \colhead{[$\MeV$]} &
\colhead{[$\MeV$]} & \colhead{[$\fme$]} & \colhead{[$\MeV$]}
}
\startdata
I   & $8.8$   & $40$ & $280$  & $0.53$ & $4.8$    & $1.5$   & $115$ & $44.3$ & $2.7$  & $90.0$ & $5.9$ & $0.42$  & $0.16$ & $0.21$ & $15.6$ & $0.21$   \\
II  & $57.7$  & $50$ & $1050$ & $0.45$ & $47.0$   & $9.6$   & $161$ & $35.7$ & $5.2$  & $72.5$ & $6.7$ & $-0.13$ & $0.94$ & $0.68$ & $10.1$ & $0.29$   \\
III & $204.0$ & $50$ & $1750$ & $0.35$ & $184.0$  & $33.9$  & $193$ & $27.6$ & $11.3$ & $56.1$ & $7.2$ & $-1.64$ & $1.40$ & $0.91$ & $12.0$ & $-2.74$  \\
IV  & $475.0$ & $40$ & $1460$ & $0.28$ & $436.0$  & $78.9$  & $183$ & $19.6$ & $31.7$ & $39.8$ & $7.3$ & $-1.00$ & $1.00$ & $0.56$ & $26.1$ & $-0.72$  \\
V   & $789.0$ & $32$ & $950$  & $0.16$ & $737.0$  & $131.0$ & $232$ & $14.4$ & $80.3$ & $29.2$ & $7.2$ & $-0.78$ & $0.49$ & $0.19$ & $90.8$ & $-0.02$  \\
\enddata
\tablenotetext{}{\textsuperscript{a}\citet{Donati2006} use different values for $R_{\rm N}$ in their calculation of $E_{\rm p}$. We choose the parameters of
\citet{Epstein1992} because they are in better agreement with the results of \citet{Negele1973}. \newline
\textsuperscript{b}We follow \citet{Epstein1992} and choose a short-range contribution that is reduced by a factor $10$. \newline
\textsuperscript{c}Note that \citet{Donati2006} use a different definition of $\xi$, resulting in a factor $1/\sqrt{6}$ instead of $1/\pi$.}
\end{deluxetable*}

\citet{Epstein1992} consider a vortex-nucleus interaction potential
\begin{equation}
	E(s) = \frac{E_{\rm s}}{(1 + s^2/R_{\rm N}^2)^4} + \frac{E_{\rm l}}{1 + s^2/R_{\rm N}^2},
  \label{eqn-potentialEB}
\end{equation}
where $s$ is the vortex-nucleus separation and $E_{\rm s}$ ($E_{\rm l}$) the short-range (long-range) contribution \citep{Epstein1988}. The potential falls off on the scale of the nuclear radius $R_{\rm N}$, corresponding to $\ell \simeq R_{\rm N}$ and
\begin{equation}
  \CR_{\rm EB} \simeq  1.4 \left(\frac{\mu} {\hbar}\right)^{\! 1/2} \left(\frac{E_{\rm p}}{\rho_{\rm s} \kappa}\right)^{\! 2}
		\frac{1}{\Delta v^{3/2}} \frac{n_l}{R_{\rm N}^{1/2}},
  	\label{eqn-REBgeneral}
\end{equation}
in agreement with the \citet{Epstein1992} scalings, and we include an appropriate numerical prefactor. Performing a more detailed analysis of the Kelvin wave excitation process and employing an interaction potential of the form~\eqref{eqn-potentialEB}, $E_{\rm p}$ is found to be a mixture of $E_{\rm s}$ and $E_{\rm l}$, with coefficients that depend on the scalings of each term with $s$. We obtain
\begin{equation}
   E_{\rm p}^2 \simeq E_{\rm s}^2 + E_{\rm l} E_{\rm s} + 0.5 E_{\rm l}^2.
    \label{eqn-energy}
\end{equation}
Note that these coefficients as well as the numerical prefactor in Equation~\eqref{eqn-REBgeneral} disagree with the results of \citet{Epstein1992}. Repeating the calculation outlined in their Appendix B and Section 3, we determine drag coefficients that are about one order of magnitude smaller than those corresponding to Equation~(3.18) in \citet{Epstein1992}. We trace the disagreement and different coefficients in Equation~\eqref{eqn-energy} back to an erroneous integration in the energy associated with the Kelvin wave excitations and/or power dissipated.\footnote{\col{More precisely, by combining the Equations (3.15) and (3.16) with (B14) in \citet{Epstein1992}, we cannot reproduce their Equation~(3.18). Note also that expression (B14) for the Fourier-transformed interaction force misses an overall factor $1/\sqrt{2}$ and the term $K^4$. These typos do however not cause the discrepancy, which instead has to originate from the integrations in Equations (3.15) or (3.16).}}

The second study of Kelvin wave dynamics adopts a different prescription for the vortex-nucleus interaction: According to \citet{Jones1992}, this process dissipates the power $p \sim \Delta E / \tau a$ per unit length, where $a$ denotes the bcc lattice constant. The drag coefficient now reads
\begin{equation}
  \CR \sim \left(\frac{\mu} {\hbar}\right)^{\! 1/2} \left(\frac{E_{\rm p}}{\rho_{\rm s} \kappa}\right)^{\! 2} \frac{1}{\Delta v^{3/2}} \frac{1}{a \ell^{5/2}}.
  \label{eqn-scalingRJ}
\end{equation}
Further, \citet{Jones1992} does not account for a long-range contribution and uses a short-range potential
\begin{equation}
	E(s) = E_{\rm p} \exp\left(-\frac{s^2}{2 \xi^2}\right),
\end{equation}
that falls off on a much larger scale, the coherence length $\xi$. The appropriate choice is now $\ell \simeq \xi$ and we find
\begin{equation}
  \CR_{\rm J} \simeq \frac{1}{2\sqrt{\pi}} \left(\frac{\mu} {\hbar}\right)^{\! 1/2} \left(\frac{E_{\rm p}}{\rho_{\rm s} \kappa}\right)^{\! 2} \frac{1}{\Delta v^{3/2}}
  \frac{1}{a \xi^{5/2}},
  \label{eqn-RJgeneral}
\end{equation}
reproducing the scalings of \citet{Jones1992} and we added his numerical prefactor.

\begin{deluxetable*}{c|ccc|ccc|cc}[t!]
\tablecaption{\col{Relative vortex-nucleus velocities and pinning forces per unit length for five crustal layers. $\Delta v$ and $f$ are calculated based on three different microscopic models: (A) expression~\eqref{eqn-energy} including $E_{\rm s,l}$ with $l \simeq R_{\rm N}$, (B) $E_{\rm p}$ with $l \simeq R_{\rm N}$, (C) $E_{\rm p}$ with $l \simeq \xi$. Additionally, estimates for the pinning forces per unit length by \citet{Seveso2016} (for $\beta = 3$) are shown. $f_{\rm S}$ ($L$ fixed) corresponds to the pinning force acting on a vortex of fixed length $L \sim 10^3 \, R_{\rm WS}$, whereas $f_{\rm S}$ ($L$ varied) allows for changes in $L$ across the inner crust due to density-dependent vortex tension. This significantly reduces $f_{\rm S}$ in domain V, because a vortex remains straight over longer distances. For more details see \citet{Seveso2016}}.
\label{tab-02}}
\tablecolumns{9}
\tablehead{
\colhead{} & \colhead{$\Delta v$ (A)} & \colhead{$\Delta v$ (B)} & \colhead{$\Delta v$ (C)} & \colhead{$f$ (A)} & \colhead{$f$ (B)} & \colhead{$f$ (C)} & \colhead{$f_{\rm S}$ ($L$ fixed)} & \colhead{$f_{\rm S}$ ($L$ varied)} \\
\colhead{} & \colhead{[$10^{4} \times$} & \colhead{[$10^{4} \times$} & \colhead{[$10^{4} \times$} & \colhead{[$10^{15} \times$} & \colhead{[$10^{15} \times$} & \colhead{[$10^{15} \times$} & \colhead{[$10^{15} \times$} & \colhead{[$10^{15} \times$} \\
\colhead{ } & \colhead{$\cm \, \se^{-1}$]} & \colhead{$\cm \, \se^{-1}$]} & \colhead{$\cm \, \se^{-1}$]} & \colhead{$\dyn \, \cm^{-1}$]} & \colhead{$\dyn \, \cm^{-1}$]} & \colhead{$\dyn \, \cm^{-1}$]} & \colhead{$\dyn \, \cm^{-1}$]} & \colhead{$\dyn \, \cm^{-1}$]}
}
\startdata
I   & $96.074$   & $39.844$ & $15.069$  & $1.53$ &  $0.63$  & $0.24$   & $0.13$ & $0.32$  \\
II  & $12.288$   &  $6.143$ &  $4.075$  & $1.91$ &  $0.96$  & $0.63$   & $0.29$ & $0.31$  \\
III &  $7.626$   & $17.829$ & $10.697$  & $4.65$ & $10.87$  & $6.52$   & $3.40$ & $8.55$  \\
IV  &  $2.700$   &  $2.749$ &  $0.769$  & $3.90$ &  $3.97$  & $1.11$   & $2.35$ & $1.84$  \\
V   &  $1.836$   &  $0.062$ &  $0.005$  & $4.48$ &  $0.15$  & $0.01$   & $0.27$ & $0.06$  \\
\enddata
\end{deluxetable*}

\col{Equations~\eqref{eqn-REBgeneral} and \eqref{eqn-RJgeneral} would suggest that the dissipation associated with Kelvin wave excitations depends sensitively on $E_{\rm p}$ and $\Delta v$. These quantities are however not independent. On small scales, vortex unpinning is initiated once the Magnus force, generated by the background superfluid, exceeds the pinning force. As these dynamics are governed by the forces acting on a vortex \textit{per unit length}, microscopic pinning interactions have to be modified to account for the finite length of vortices. In the inner crust, these structures remain straight over a distance $L \sim 10^3 \, R_{\rm WS}$ \citep{Seveso2016} and interact with many randomly orientated nuclei. By geometrically averaging over this mesoscopic scale, \citet{Seveso2016} determine a decrease in the pinning force per unit length by about two orders of magnitude in agreement with \citet{Jones1990, Jones1992}. We include this by accounting for a constant \textit{reduction factor} $\delta \approx 10^{-2}$ and introduce effective pinning energies $E_{\rm p} \to E_{\rm p} \delta$. By balancing the Magnus force and pinning force per unit length, the critical velocity lag $\Delta v_{\rm cr}$ between a vortex and the background superfluid flow at which unpinning takes place can thus be related to the microscopic parameters characterizing the pinning interaction.} \col{The local relative velocity $\Delta v$ between a free vortex and the nuclear lattice is typically of the same order as $\Delta v_{\rm cr}$ and we estimate
\begin{equation}
  \Delta v \simeq \Delta v_{\rm cr} = \frac{f}{\rho_{\rm s} \kappa} \sim \frac{E_{\rm p} \delta}{l a \rho_{\rm s} \kappa}.
    \label{eqn-forcebalance}
\end{equation}
Substituting this into Equations~\eqref{eqn-REBgeneral} and \eqref{eqn-RJgeneral} gives
\begin{align}
  \CR_{\rm EB} &\simeq  2.8 \left(\frac{\mu} {\hbar}\right)^{\! 1/2} \left(\frac{E_{\rm p} \delta}{\rho_{\rm s} \kappa}\right)^{\! 1/2}
    \frac{R_{\rm N}}{a^{3/2}},
		\label{eqn-dragfinalEB} \\[1.2ex]
  \CR_{\rm J} &\simeq \frac{1}{2\sqrt{\pi}} \left(\frac{\mu} {\hbar}\right)^{\! 1/2} \left(\frac{E_{\rm p}\delta}{\rho_{\rm s} \kappa}\right)^{\! 1/2}
    \frac{a^{1/2}} {\xi}.
		\label{eqn-dragfinalJ}
\end{align}
The two expressions differ by
\begin{equation}
  \frac{\CR_{\rm J}}{\CR_{\rm EB}} \simeq 0.1 \, \frac{a^2}{R_{\rm N} \xi}.
\end{equation}}

In the next section, we calculate these coefficients for a realistic crust model, and show that the different choices for the vortex-nucleus interaction affect the strength of the crustal mutual friction.


\begin{figure}[t!]
\centering
\includegraphics[width = 0.47\textwidth]{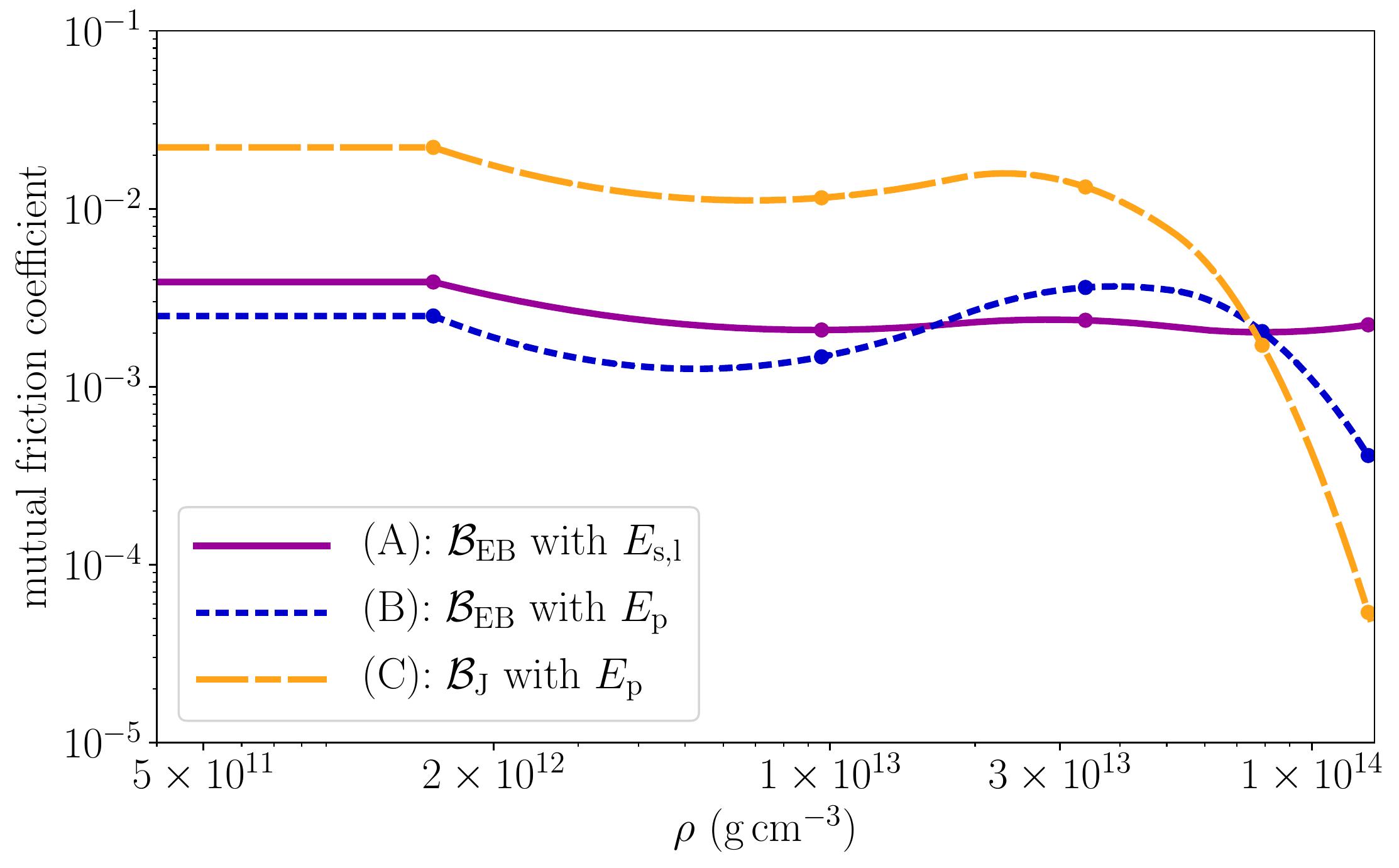}
\includegraphics[width = 0.47\textwidth]{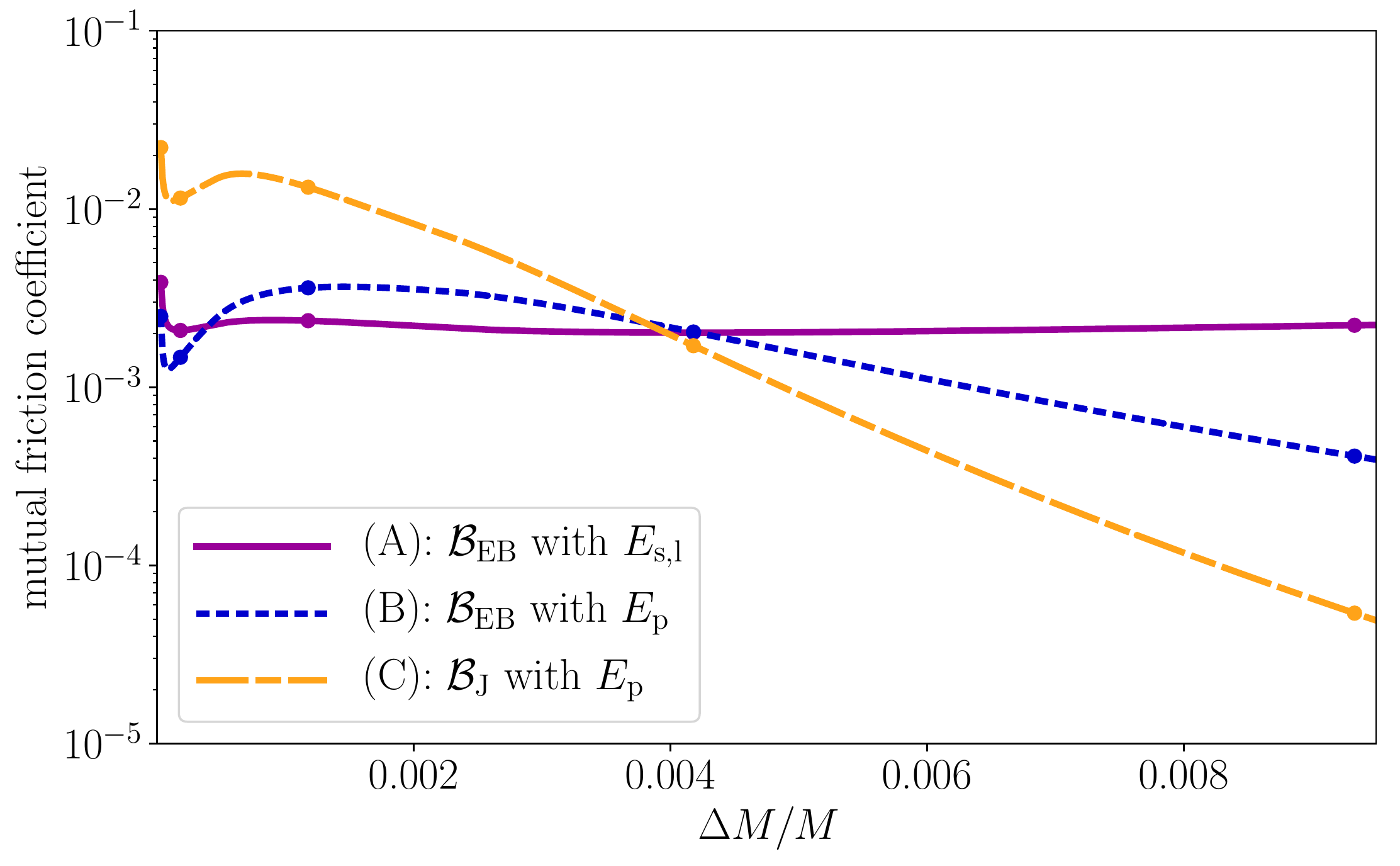}
\caption{Mutual friction strength associated with Kelvin wave excitation as a function of mass density $\rho$ (upper panel) and overlying relative mass fraction $\Delta M / M$ (lower panel). Based upon different assumptions on the microscopic vortex-nucleus interaction, $\CB$ is calculated in three different ways.}
\label{fig:mf_crust}
\end{figure}

\section{Density-dependent coupling for a realistic crust model}
\label{sec:microphysics}

Microscopic parameters for five inner crustal regions are summarized in Table~\ref{tab-01}. These are based on the EoS of \citet{Negele1973}, calculated in the Wigner-Seitz approximation. Adopting this ground-state composition, several authors have studied the vortex-nucleus interaction by analyzing the energy gain/loss of superimposing a vortex and a single lattice site. Table~\ref{tab-01} shows estimates for $R_{\rm N}$, $E_{\rm s}$ and $E_{\rm l}$ determined by \citet{Epstein1992} using a Ginzburg-Landau approximation. Whereas $E_{\rm s}$ changes sign and can be repulsive or attractive, $E_{\rm l}$ is of hydrodynamical origin (related to the change in the superfluid's kinetic energy due to the nucleus as a result of the Bernoulli effect) and always repulsive \citep{Shaham1980, Epstein1988}, leading to different pinning geometries. More recently, \citet{Donati2006} have studied the pinning problem within a semi-classical model. Their estimates for $\Delta$, $\xi$ and $E_{\rm p}$ are also shown in Table~\ref{tab-01}. Due to the competition of the superfluid's internal, kinetic and condensation energy \citep{Donati2004}, they also obtain distinct configurations: \textit{interstitial pinning} at lower densities ($E_{\rm p} > 0 $ implies vortex-nucleus repulsion) and \textit{nuclear pinning} at high densities (vortex-nucleus attraction due to $E_{\rm p} < 0$). Additionally, \citet{Donati2006} argue that the pinning strength decreases significantly towards the crust-core boundary due to \textit{collective pinning}: each vortex contains several nuclei because the Wigner-Seitz radius $R_{\rm WS}$ is smaller than $\xi$, effectively weakening the interaction.

\col{Combining these microscopic parameters with Equation~\eqref{eqn-forcebalance}, the relative vortex-nucleus velocities and pinning forces can be evaluated for the five crustal layers. To assess the impact of different assumptions about the pinning interaction, and allow a comparison between the formalisms of \citet{Epstein1992} and \cite{Jones1992}, we calculate $\Delta v$ and $f$ for three distinct cases: (A) the full expression~\eqref{eqn-energy} including $E_{\rm s,l}$ together with $l \simeq R_{\rm N}$, (B) $E_{\rm p}$ with $l \simeq R_{\rm N}$, and (C) $E_{\rm p}$ with $l \simeq \xi$. The resulting estimates are given in Table~\ref{tab-02}. We observe that $\Delta v$ changes significantly between the microscopic models and can differ by up to two orders of magnitude from the macroscopically averaged velocity lag $\Delta v_{\rm av}$ (see also \citealt{Gugercinoglu2016}).\footnote{\col{For typical Vela pulsar parameters we can estimate an averaged velocity difference via $\Delta v_{\rm av} \simeq \Delta \Omega_{\rm crit} R \simeq |\dot{\Omega}_{\rm crust}| t_{\rm glitch} R \sim  10^4\, \cm \, \se^{-1}$, with a stellar radius $R \sim 10 \, \km$, an observed pre-glitch spin-down rate $|\dot{\Omega}_{\rm crust}| \sim 10^{-10} \, \rad \, \se^{-2}$ \citep{Dodson2007} and an inter-glitch time $t_{\rm glitch} \sim 3 \, \yr$.}}
Note that according to \citet{Jones1992}, Kelvin wave dissipation is only effective if $\Delta v \gtrsim 10^2 \, \cm \, \se^{-1}$. For lower relative velocities, the energy loss proceeds via the excitations of lattice phonons, which is much weaker \citep{Jones1990, Jones1992}. One velocity estimate in Table~\ref{tab-02} ($\Delta v \approx 50 \, \cm \, \se^{-1}$ for case (C) in zone V) drops slightly below this limit, suggesting that strong Kelvin wave drag might be replaced by weaker phonon drag close to the crust-core interface. Since all but one $\Delta v$-estimates indicate a Kelvin-wave dominated regime, we focus on the friction coefficients presented in Section ~\ref{sec:coupling} and postpone an analysis of the impact of phonon-related dissipation to future work. Table~\ref{tab-02} further provides the pinning forces per unit length given by \citet{Seveso2016}. Comparison with our estimates shows that accounting for the geometric average over a mesoscopic vortex segment via a reduction factor $\delta$ gives reasonable agreement and discrepancies can be attributed to different microscopic treatments of the pinning energy and the corresponding interaction length scales.}

\begin{figure*}[t!]
\centering
\gridline{
\includegraphics[height = 0.29\textwidth]{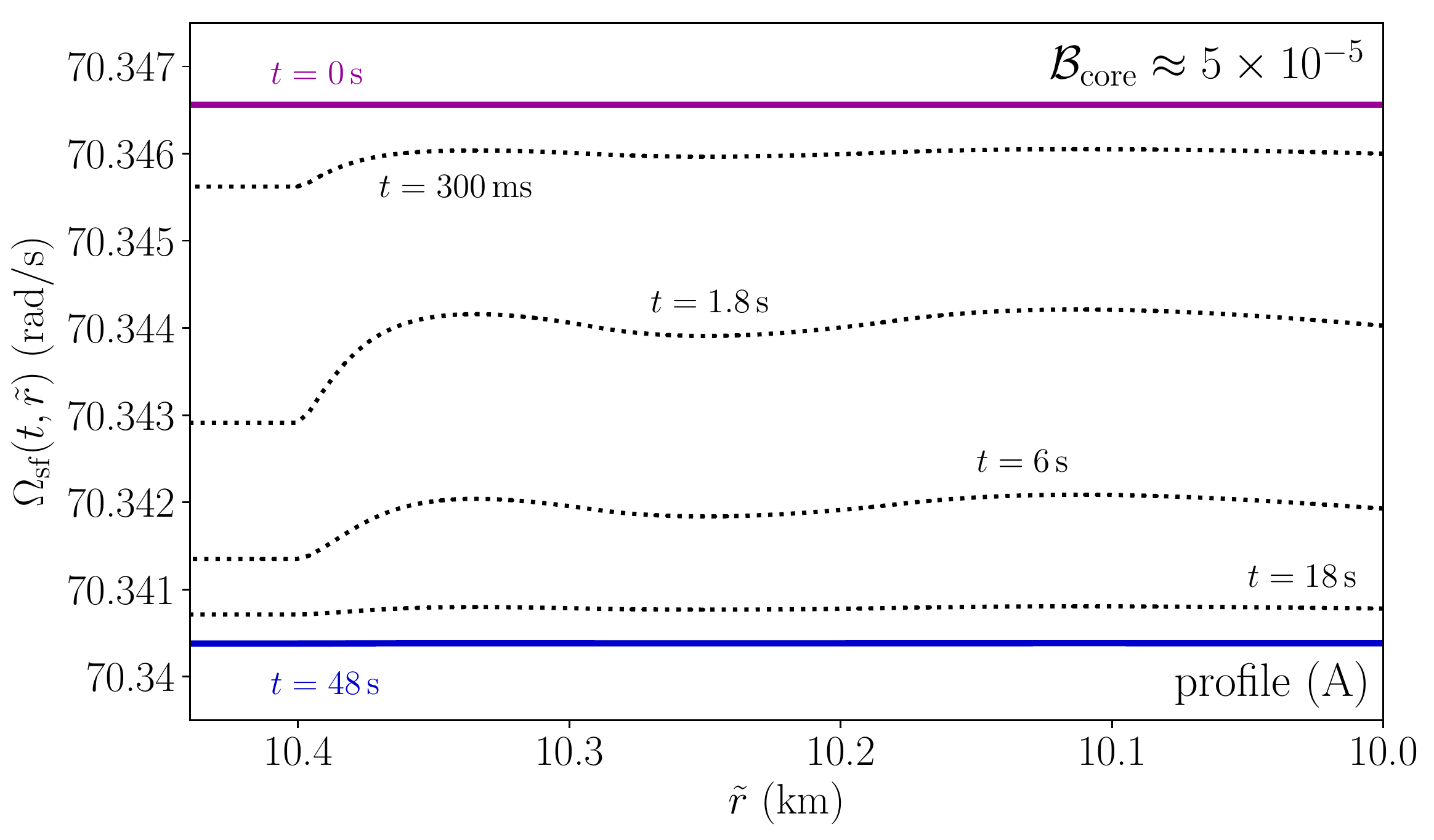}
\includegraphics[height = 0.29\textwidth]{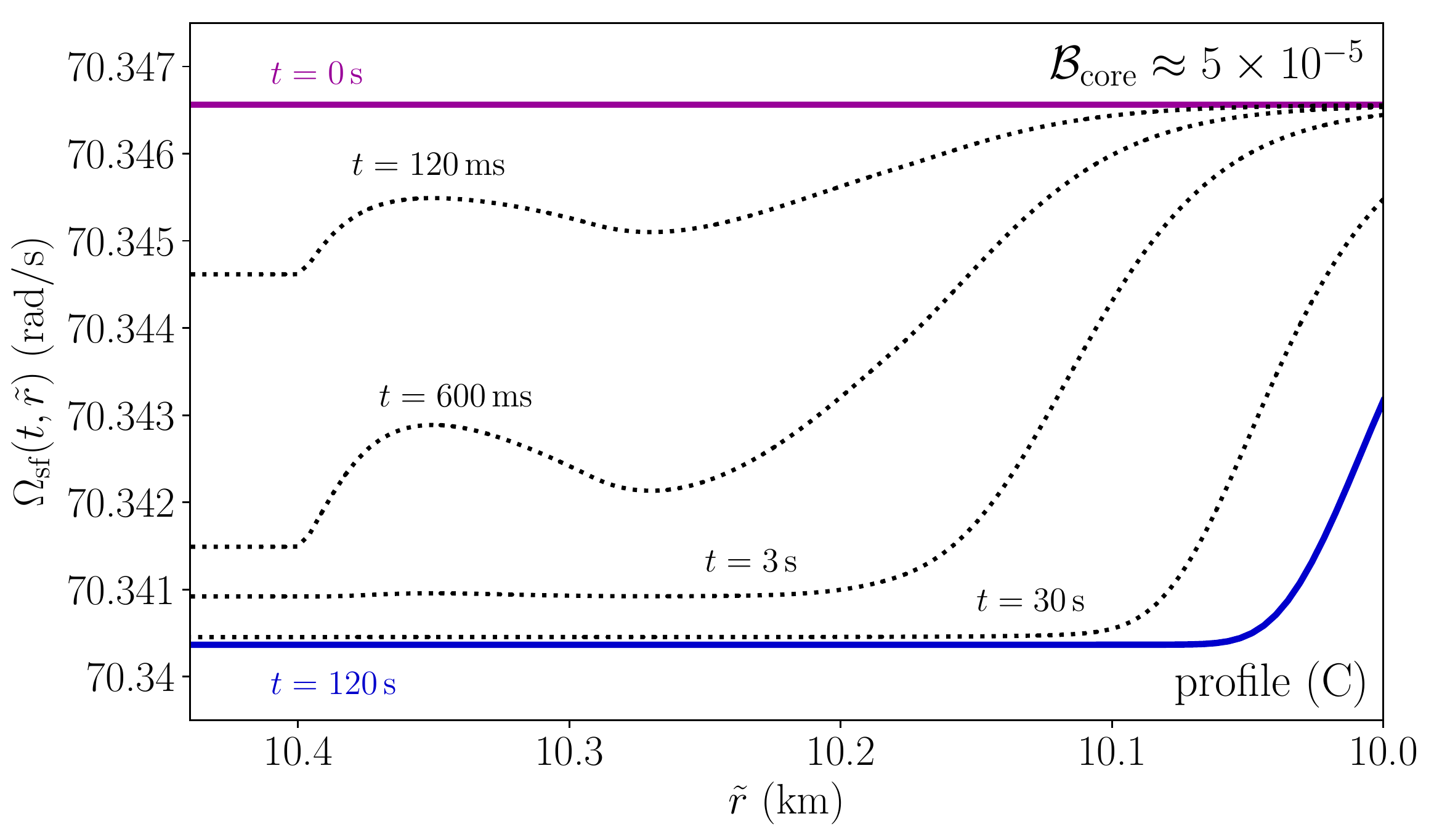}}
\vspace{-0.4cm}
\gridline{
\includegraphics[height = 0.29\textwidth]{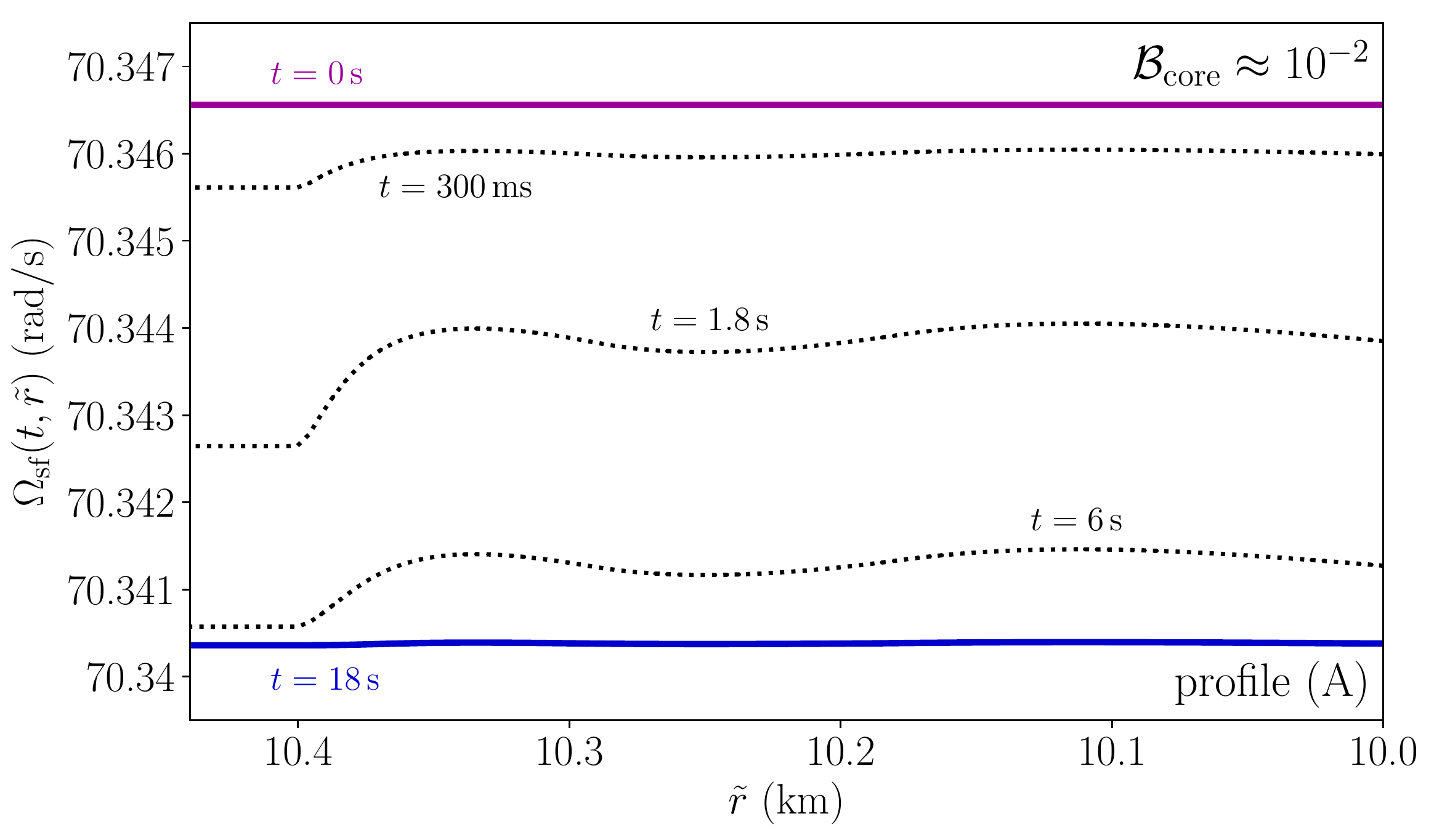}
\includegraphics[height = 0.29\textwidth]{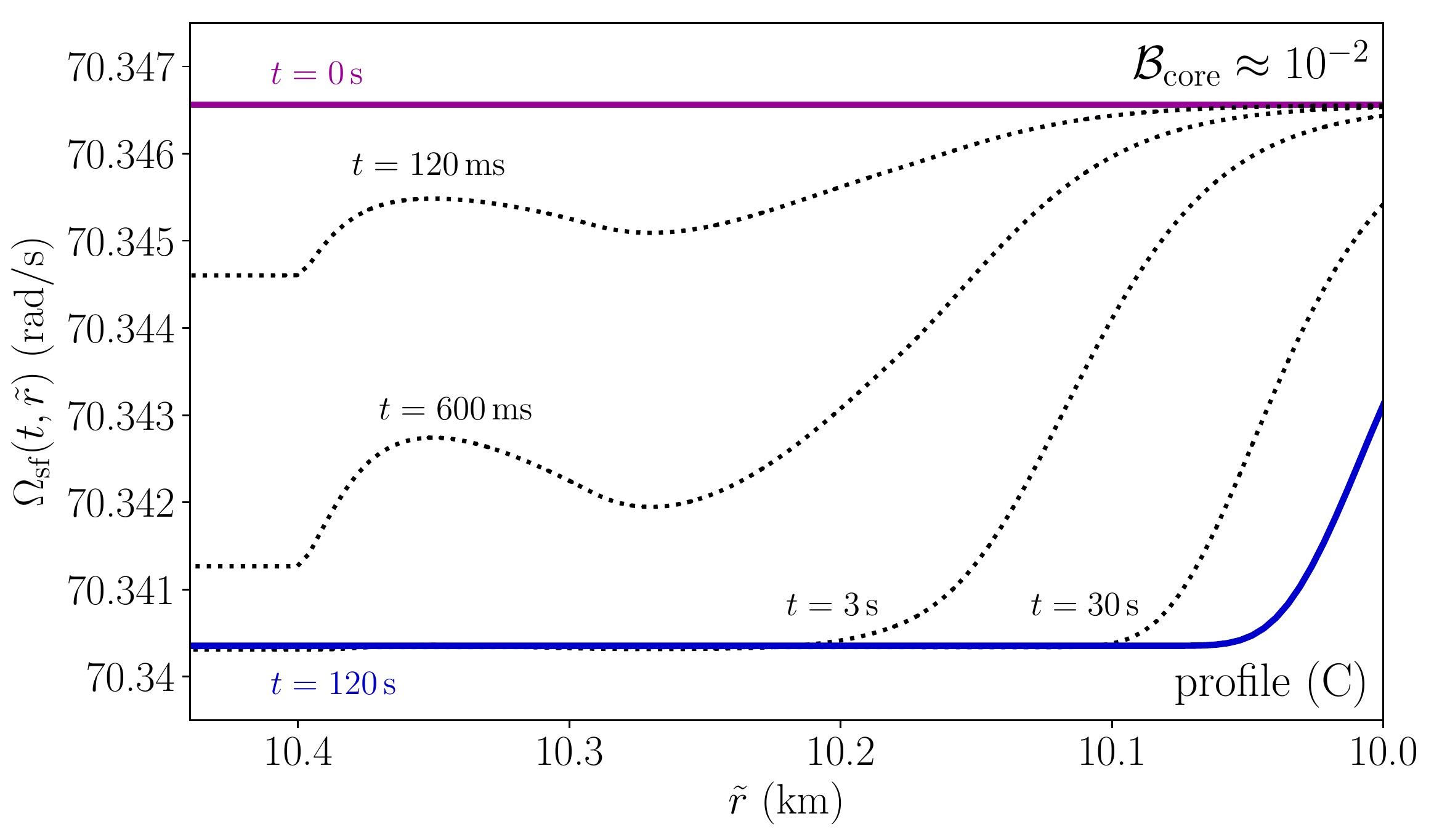}}
\vspace{-0.3cm}
\caption{\col{$\Omega_{\rm sf}$ as function of radius and time. Results are shown between neutron drip and crust-core interface and calculated for drag profiles (A) (left panels) and (C) (right panels) together with $\CB_{\rm core} \approx 5 \times 10^{-5}$ (top panels) and $\CB_{\rm core} \approx 10^{-2}$ (bottom panels). Purple lines mark the superfluid's initial rotation, while blue lines represent the new steady state in the case of model (A) and the $\Omega_{\rm sf}$ profile at $120 \, \se$ for model (C), where angular momentum transfer has not been completed. Black, dotted lines show $\Omega_{\rm sf}$ at different times.}}
\label{fig:time_evolution}
\end{figure*}

To evaluate the drag coefficients of Section~\ref{sec:coupling}, we require the effective mass. Since the vortex tension can be approximated with $T \simeq - \rho_{\rm s} \kappa^2 \ln (k \xi) / 4 \pi$ \citep{Sonin1987}, we obtain $\mu(k) \simeq - 2 m_{\rm u} /\ln k \xi$.\footnote{Note there is a missing factor of $2\pi$ in the expressions for $\mu$ in Section 3 of \citet{Epstein1992} and in \citet{Link2003}, which has propagated in the literature. In their notation, $\mu(k) \equiv m_{\rm u}/\pi \lambda(k)$, $\lambda(k)$ is missing a factor of $1/2\pi$.} To simplify the calculation, \col{we neglect the density-dependence of $\mu$ and evaluate the logarithmic factor at a characteristic wave number $\col{k_* \approx 1.5 \times 10^{-3} \, \fme^{-1}}$ and typical coherence length $\xi \approx 12 \fme$, which gives $\col{\mu \simeq \mu(k_*) \approx m_{\rm u} /2}$. The error introduced by keeping $\mu$ constant is small due to the weak dependence on $k$ and $\xi$.} Using fiducial values for domain III, Equations~\eqref{eqn-dragfinalEB} and \eqref{eqn-dragfinalJ} yield
\begin{align}
    \col{ \CR_{\rm EB}} &\approx 3.5 \times 10^{-3} \left( \frac{|E_{\rm p}|}{3 \, \MeV} \right)^{\!1/2}
    \left( \frac{\delta}{10^{-2}} \right)^{\! 1/2} \nonumber\\
    &\times \left( \frac{2 \times 10^{-2} \, \fme^{-3}}{n_{\rm s}} \right)^{\! 1/2}  \left( \frac{56 \, \fme}{a} \right)^{\! 3/2} \! \left( \frac{R_{\rm N}}{7 \, \fme} \right)
\end{align}
and
\begin{align}
    \col{\CR_{\rm J} }&\approx 1.3 \times 10^{-2} \left( \frac{|E_{\rm p}|}{3 \, \MeV} \right)^{\!2}
      \left( \frac{\delta}{10^{-2}} \right)^{\!2} \nonumber \\
    & \times \left( \frac{2 \times 10^{-2} \, \fme^{-3}}{n_{\rm s}} \right)^{\! 2}  \left( \frac{a}{56 \, \fme} \right)^{\! 1/2}
      \left( \frac{12 \, \fme}{\xi} \right).
\end{align}
\col{As before, we compare the assumptions about the microphysics in the formalisms of \cite{Epstein1992} and \cite{Jones1992} (long- plus short-range interactions versus short-range interaction only) by determining the friction coefficient in three different ways: (A) $\CR_{\rm EB}$ calculated with the full expression~\eqref{eqn-energy} including $E_{\rm s,l}$, (B) $\CR_{\rm EB}$ calculated with $E_{\rm p}$ only, and (C) $\CR_{\rm J}$ calculated with $E_{\rm p}$. We refer to these cases with the labels (A), (B), and (C).}

The resulting profiles of $\CB$ are illustrated in Figure~\ref{fig:mf_crust}.\footnote{A Jupyter Notebook to reproduce plots and results is publicly available at \url{https://github.com/vanessagraber/glitchrises}.} We employ a spline function to interpolate results for the five domains from neutron drip at $\rho_{\rm D} \approx 4.0 \times 10^{11} \, \g \, \cm^{-3}$ to the crust-core interface at $\rho_{\rm cc} \approx 1.3 \times 10^{14} \, \g \, \cm^{-3}$.
The fit gives unphysical results when extrapolating below $1.5 \times 10^{12} \, \g \, \cm^{-3}$; instead we take $\CB$ constant for simplicity. We observe that mutual friction varies strongly with density and differs significantly close to the crust-core interface. This region carries the majority of the superfluid's mass as shown in the bottom panel of Figure~\ref{fig:mf_crust}, where $\CB$ is given as a function of relative mass fraction $\Delta M / M$. Here, $\Delta M$ denotes the overlying mass and $M$ the total mass taken to be $M \approx 1.41 M_{\odot}$ ($M_{\odot}$ is the solar mass) in our model. The range of $\CB$ suggests different post-glitch behavior.


\section{Glitch rise modeling}
\label{sec:toymodel}

To analyze the effects of density-dependent friction on the post-glitch response, we use a simple time-dependent three-component model to determine the shape of the glitch rise. The star is decomposed into a crust neutron superfluid, core neutron superfluid and a non-superfluid `crust' component, representing the nuclear lattice and tightly coupled charged conglomerate in the core \citep{Easson1979a}. Angular velocities and moments of inertia are denoted by $\Omega_{\rm x}$ and $I_{\rm x}$ with $x \in \{ {\rm sf, core, crust} \}$, respectively. We also assume the crust and core components to be rigidly rotating and incorporate crust-core coupling by assigning a constant mutual friction coefficient $\CB_{\rm core}$. To account for uncertainties in our understanding of the underlying mechanism, we determine the glitch rise for two fiducial values. If the dynamics are dominated by the scattering of electrons off magnetized vortices \citep{Alpar1984}, $\CB_{\rm core} \approx 5 \times 10^{-5}$ \citep{Andersson2006b} is a suitable choice. Stronger friction could be present if the crust-core coupling is mediated by the interactions between vortices and superconducting fluxtubes \citep{Link2003, Sidery2009}, provided that the core protons form a type-II state \citep{Baym1969}. To study this possibility, we  \col{follow \citet{Link2003} and \citet{Haskell2014} and assume that Kelvin waves are excited along the vortices as they cut through fluxtubes. The formalism discussed in Section~\ref{sec:coupling} can be directly translated provided that the microphysics are adjusted. The vortex-fluxtube interaction is predominantly magnetic, resulting in pinning energies on the order of $E_{\rm p} \approx 5 \, \MeV$ \citep{Link2003} active over a length scale $l \simeq \lambda_* \approx 100 \, \fme$, i.e. the London penetration depth. See e.g. \citet{Graber2017} for typical length scale estimates. For a fiducial magnetic field strength of $B \approx 10^{12} \, \G$, the inter-fluxtube distance is $d_{\rm ft} \approx 10^3 \, \fme$, leading to pinning forces per unit length of $f \approx 8 \times 10^{15} \, \dyn \, \cm^{-1}$. As for the crust, a local force balance provides the means to estimate the relative vortex-fluxtube velocity to $\Delta v \approx 4 \times 10^4 \, \cm \, \se^{-1}$ for $\rho_{\rm s} \approx 10^{14} \, \g \, \cm^{-3}$. For an effective mass of $\mu \approx m_u /2$, this subsequently gives $\CB_{\rm core}  \simeq \CR_{\rm core}\approx 10^{-2}$.}

\begin{figure*}[t!]
\centering
\gridline{
\includegraphics[height = 0.29\textwidth]{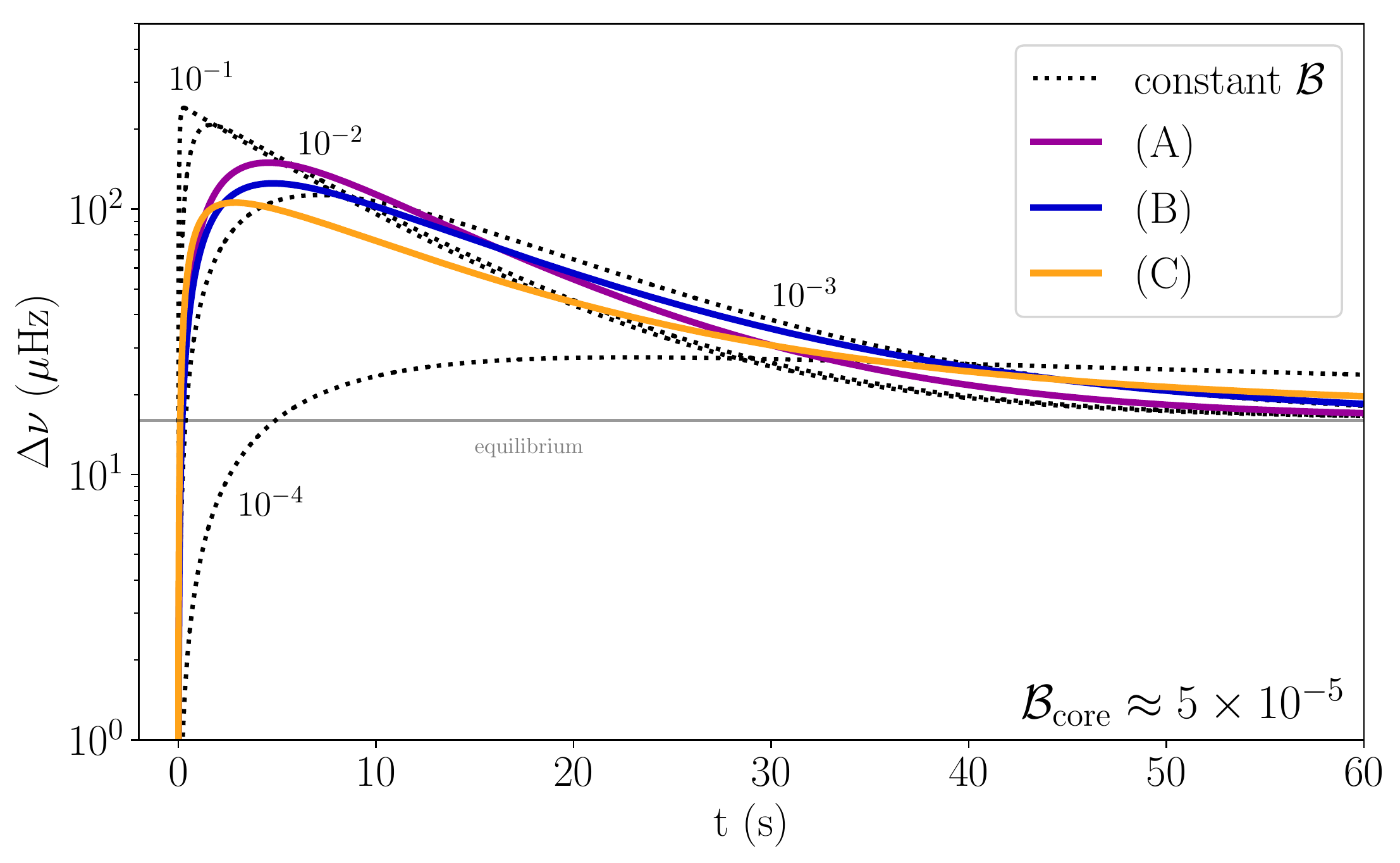}
\includegraphics[height = 0.29\textwidth]{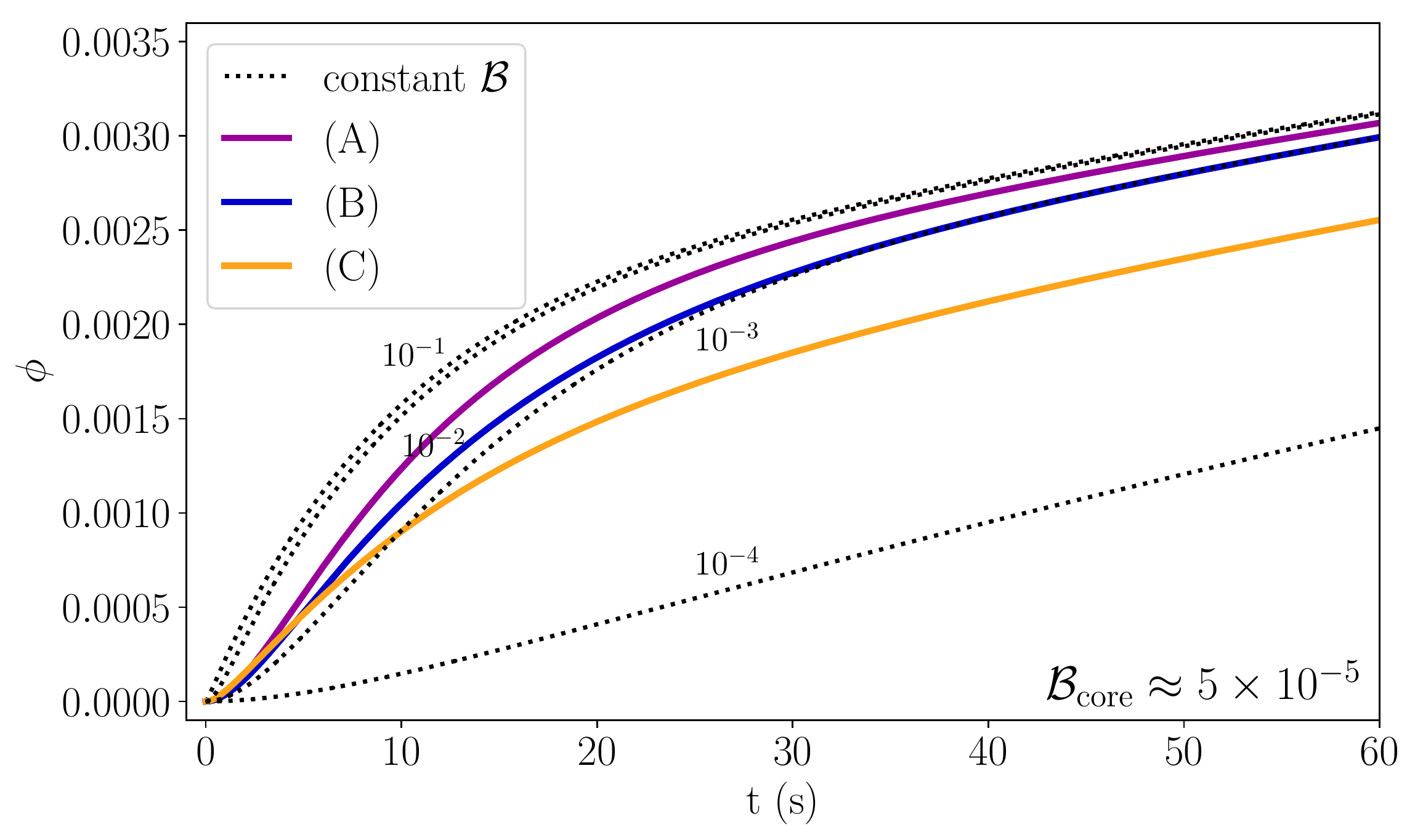}}
\vspace{-0.4cm}
\gridline{
\includegraphics[height = 0.29\textwidth]{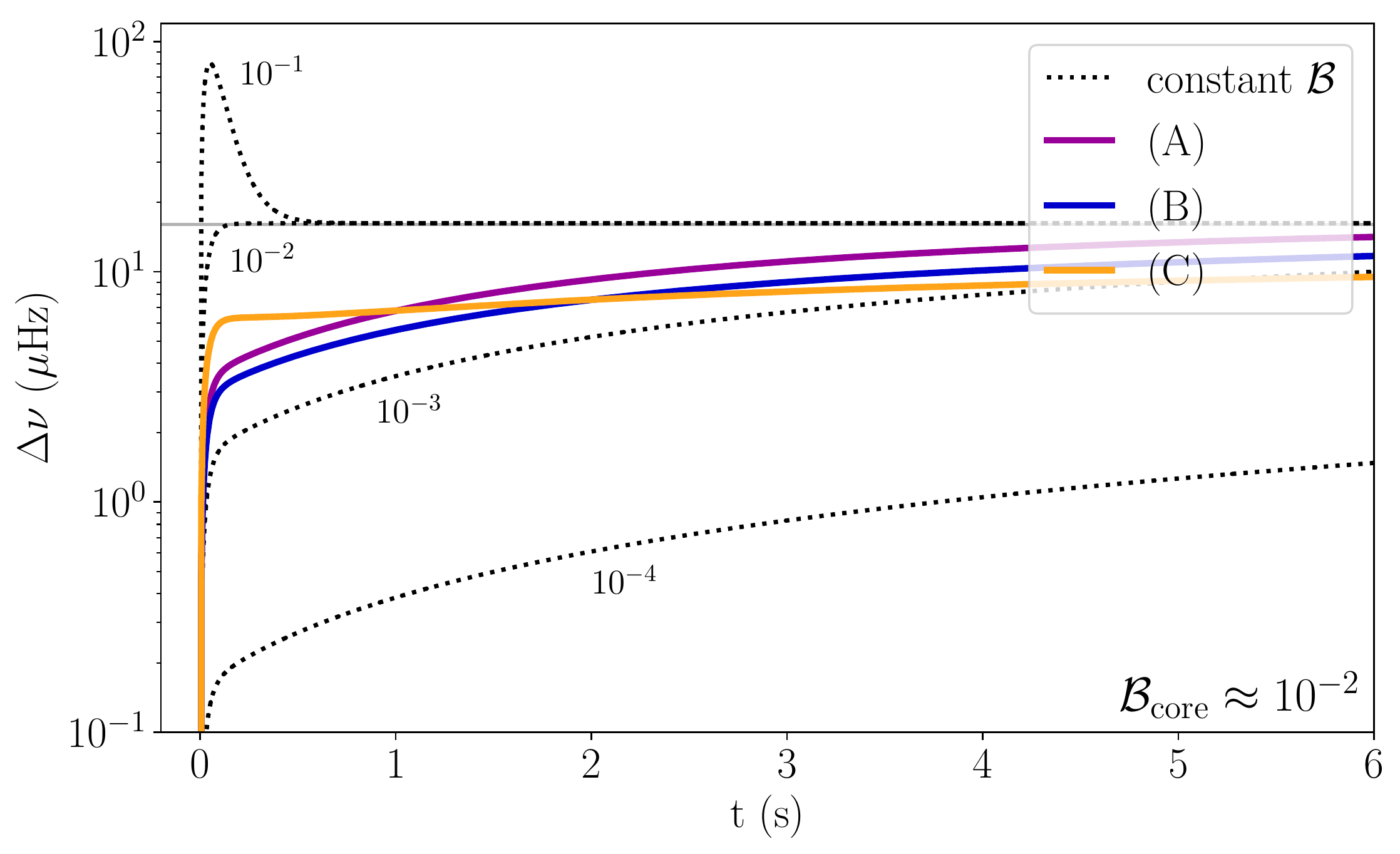}
\includegraphics[height = 0.29\textwidth]{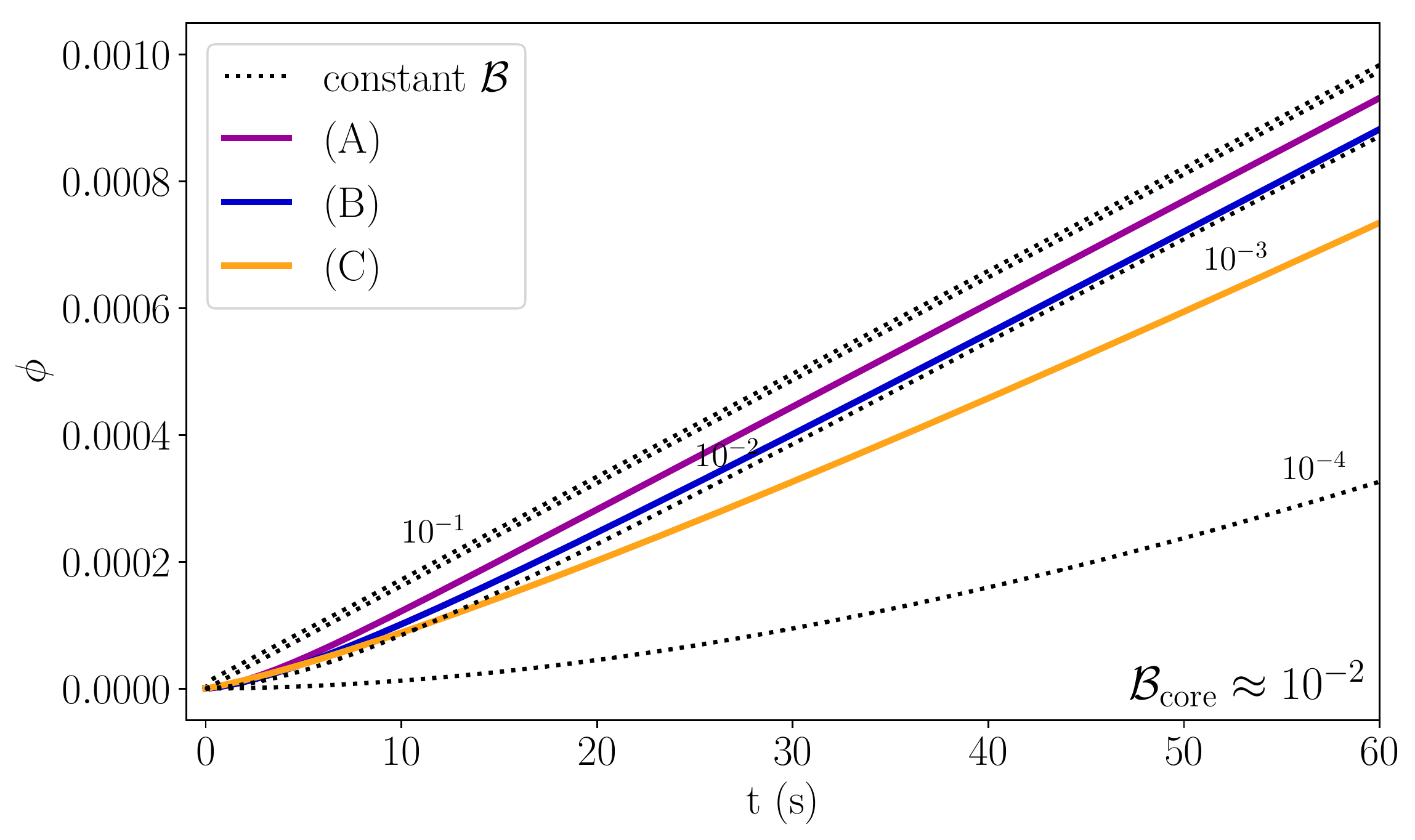}}
\vspace{-0.3cm}
\caption{Change in crustal frequency $\Delta \nu(t) = [\col{\Omega_{\rm crust}(t) - \Omega_{\rm crust}(0)}]/2\pi$ and phase shift $\phi = \int \Delta \nu \, {\rm d}t$ with time. Glitch rises are computed for three density-dependent and four constant crustal mutual friction coefficients together with $\CB_{\rm core} \approx 5 \times 10^{-5}$ (top panels) and $\CB_{\rm core} \approx 10^{-2}$ (bottom panels).
Note that we have zoomed in on the $\Delta \nu$ plot for the strong crust-core coupling scenario to show the initial post-glitch behavior.}
\label{fig:delta_nu}
\end{figure*}

Generalizing the results of \citet{Haskell2015} to three components as well as neglecting entrainment, the equations of motion read
\begin{align}
\dot{\Omega}_{\rm sf} &= \CB \left[ 2 \Omega_{\rm sf} + \tilde{r} \, \frac{\partial \Omega_{\rm sf}}{\partial \tilde{r}} \right] (\Omega_{\rm crust} - \Omega_{\rm sf}),
  \label{eqn-EOMsf} \\[1.3ex]
\dot{\Omega}_{\rm core} &= 2 \CB_ {\rm core}  \Omega_{\rm core} \, (\Omega_{\rm crust} - \Omega_{\rm core}),
\label{eqn-EOMcore}
\\[1.3ex]
\dot{\Omega}_{\rm crust} &= -  \col{\frac{N_{\rm ext}}{I_{\rm crust}}} - \frac{I_{\rm core}}{I_{\rm crust}} \, \dot{\Omega}_{\rm core}
- \frac{ \int \rho \tilde{r}^2 \dot{\Omega}_{\rm sf} \, {\rm d} V }{I_{\rm crust}},
\label{eqn-EOMcrust}
\end{align}
where \col{$N_{\rm ext}$ is the external spin-down torque}, $\tilde{r}$ the cylindrical radius and the integral is performed over the inner crust. For simplicity, a cylindrical geometry is used: We solve the problem in the equatorial plane and rescale the results so that the total crustal moment of inertia in cylindrical coordinates matches that in spherical ones. We assume a total moment of inertia of $I_{\rm tot} \approx 0.35\, M R^2$ \citep{Lattimer2001}, with the core neutrons and charged particles constituting $95\%$ and $5\%$ of the core's moment of inertia, respectively. In order to relate $\rho$ and $\tilde{r}$ in the crust, we integrate the TOV equations assuming a core radius and mass of $10\, {\rm km}$ and $1.4\, M_\odot$. For consistency, we consider the \citet{Negele1973} EoS for the inner crust. As this EoS does not apply in the outer crust, we take the pressure below neutron drip to be dominated by relativistic electrons with $Y_e \approx 0.4$.

Using initial conditions which are typical for the Vela pulsar, i.e. $\Omega_{\rm crust, \, core}(0) \approx 70.34 \, \rad \, \se^{-1}$ and \col{$\Delta \Omega_{\rm crit} \equiv \Omega_{\rm sf}(0) - \Omega_{\rm crust}(0) \approx 6.3 \times 10^{-3} \, \rad \, \se^{-1}$ to allow a comparison with the glitch observations of \citet{Palfreyman2018}}, we evolve the equations of motion \eqref{eqn-EOMsf}-\eqref{eqn-EOMcrust} for $120 \, \se$ to encompass observational constraints on the spin-up timescale \citep{Dodson2002, Dodson2007, Palfreyman2018}. \col{Note that our crustal mutual friction profiles can vary in space but remain constant over time. We further ignore the external spin-down torque when integrating the equations of motion, since it has negligible effect on the short-term post-glitch response within $120 \, \se$.}

As illustrated by the characteristic shape of $\Omega_{\rm sf}(t, \tilde{r})$ in Figure~\ref{fig:time_evolution}, the superfluid's differential rotation is mainly driven by the $\CB(\tilde{r})$-dependence.\footnote{Only for initial lags much larger than given above, does the derivative in Equation~\eqref{eqn-EOMsf} affect our results by steepening the $\Omega_{\rm sf}$ profile. This is due to the resemblance of Equation~\eqref{eqn-EOMsf} with Burgers equation as recently noted by \citet{Khomenko2018b}.} \col{We show results for the drag profiles (A) and (C), strongest and weakest at high densities, respectively. For case (A), we observe that the superfluid starts to couple within $\sim 100 \, \ms$, eventually transferring all its excess angular momentum and spinning down to a new steady state, where all three components are corotating. The superfluid's evolution looks qualitatively similar for the other two drag profiles, albeit exhibiting stronger differential rotation since they cover a larger range of mutual friction strengths than the profile (A). Moreover, for cases (B) and (C) the bottom of the crust recouples on longer timescales as $\CB$ is weaker in this region. Note that the angular momentum transfer from the superfluid to the crust component is not completed within $120 \, \se$ for model (C). Figure~\ref{fig:time_evolution} also indicates that for model (A), a stronger $\CB_{\rm core}$ causes the superfluid to reach the equilibrium faster. The impact of different core couplings on the superfluid's differential rotation is less pronounced for cases (B) and (C).}

To illustrate the effects of density-dependent crustal profiles and a variable crust-core coupling strength on observables, we compute the change in crustal frequency with time. For comparison, $\Delta \nu$ is also determined for four constant coefficients, $\CB \approx 10^{-1}, 10^{-2}, 10^{-3}, 10^{-4}$. Results for $\Delta \nu(t)$ are shown in the left panels of Figure~\ref{fig:delta_nu}, highlighting \col{the effect of the relative strength between the crust and the core couplings: For $\CB_{\rm core} \approx 5 \times 10^{-5}$, angular momentum transfer from the superfluid to the crust is very effective and acting on timescales shorter than the crust-core coupling timescale $\tau_{cc} \approx 7.5 \, \se$. This causes the crust's rotation frequency to increase above the asymptotic value, generating a characteristic `overshoot'. As soon as crust and core recouple, $\Delta \nu$ decreases and eventually approaches the new steady state. For the stronger scenario $\CB_{\rm core} \approx 10^{-2}$, crust-core coupling proceeds on $\tau_{cc} \approx 37.4 \, \ms$, which is faster than the density-dependent crust couplings. Thus, the superfluid transfers angular momentum to the combined crust-core system, resulting in a slower, monotonic rise of $\Delta \nu$. Note that as illustrated in the top right panel of Figure~\ref{fig:delta_nu}, the onset of crust-core coupling in the case $\CB_{\rm core} \approx 5 \times 10^{-5}$ is clearly visible as a break in the phase shift $\phi$, accumulating after the glitch. For $\CB_{\rm core} \approx 10^{-2}$, however, the break in phase shift $\phi$ moves to the left and becomes basically invisible. As we will illustrate in the next section, observing such a feature in $\phi$ could provide important information about the frictional processes in the core.}

\begin{figure*}[t!]
\centering
\includegraphics[height = 0.29\textwidth]{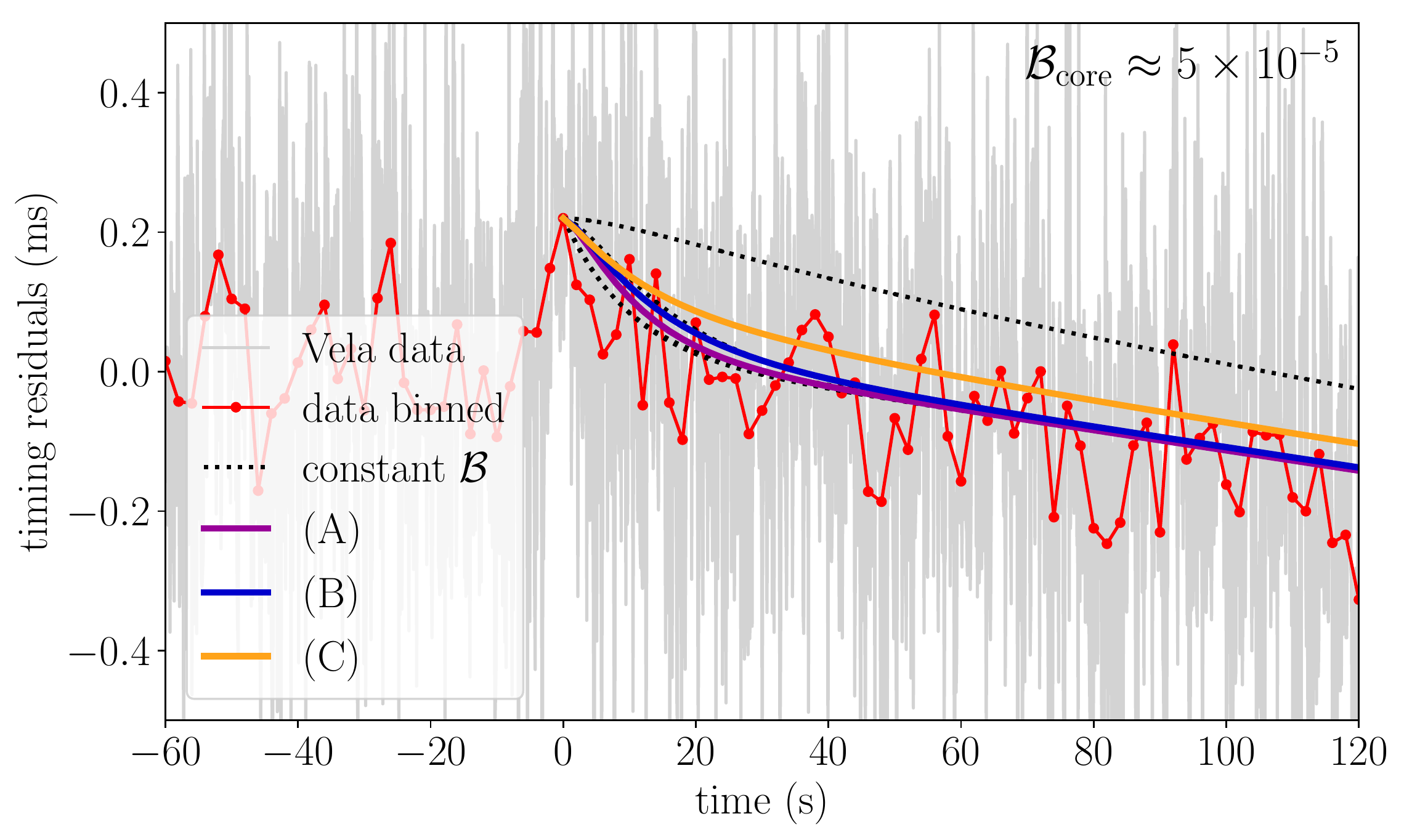}
\includegraphics[height = 0.29\textwidth]{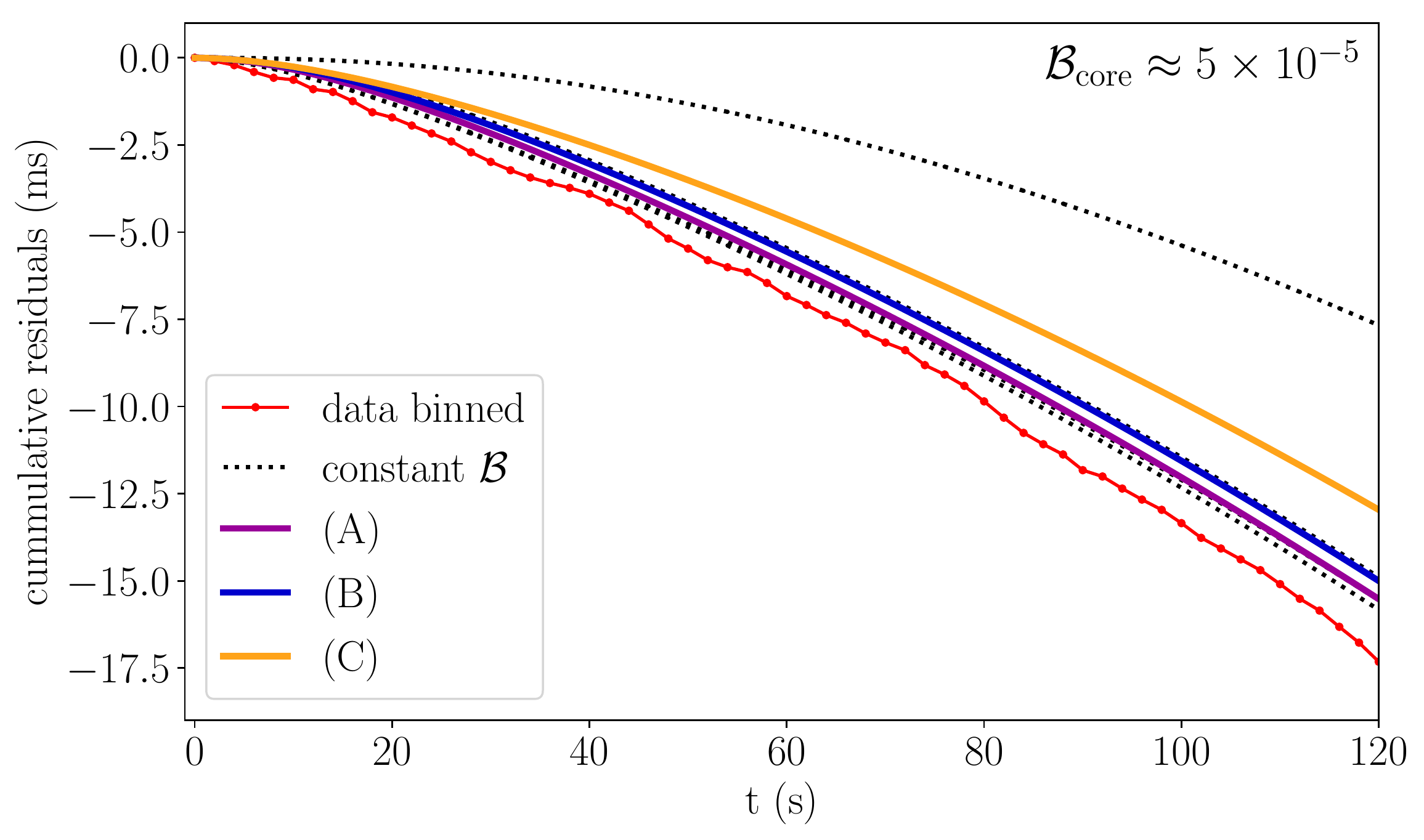}
\caption{Comparison between theoretical predictions for three density-dependent and four constant crustal mutual friction coefficients $\CB \approx 10^{-1}, 10^{-2}, 10^{-3}, 10^{-4}$ together with $\CB_{\rm core} \approx 5 \times 10^{-5}$ and observations of the 2016 Vela pulsar glitch. (Left) The original timing residuals (in milliseconds) from \citet{Palfreyman2018} (gray solid line) are centered around the glitch epoch $t_g = 57734.4849906 \,$MJD. Red points (connected by a solid red line to guide the eye) show the data averaged into $2 \, \se$ bins. Model residuals are calculated via $- 2 \pi \phi / \Omega_{\rm crust} (0)$ and shifted by $\Delta t \approx 0.22 \, \ms$ at $t = 0$. (Right) Cumulative timing residuals starting at the time of the glitch. A shift $\Delta t$ has been subtracted from the binned data points.}
\label{fig:data_comparison}
\end{figure*}

\col{Assuming that the superfluid reservoir is completely depleted, angular momentum conservation dictates for the equilibrium lag}
\begin{equation}
  \Delta \nu_{\rm equi} = \frac{I_{\rm sf}}{\col{I_{\rm total}}} \, \frac{\Delta \Omega_{\rm crit}}{2 \pi} \approx 16.0 \, \mu \Hz,
\end{equation}
in agreement with the glitch step size determined by \citet{Palfreyman2018} for the 2016 Vela pulsar glitch. Details of the approach to steady-state depend crucially on the strength of $\CB$ close to the crust-core interface. \col{Note that for profile (C), weakest at high densities, the crust does not reach the new equilibrium within the $120 \, \se$ integration window and additional time would be required for the superfluid to transfer all its angular momentum (also illustrated in the right panels of Figure~\ref{fig:time_evolution}). Such a slow recovery could in principle be misinterpreted as the new spin-equilibrium, leading to incorrect initial conditions or moment of inertia estimates.}


\section{Data comparison}
\label{sec:datacomparison}

The first pulse-to-pulse glitch observation was recently published by \citet{Palfreyman2018} for a glitch that occurred on 2016 December 12 in the Vela pulsar. To test the potential of our model as a tool to constrain microphysics, a preliminary comparison between the predictions and the new data is presented. Because various processes introduce noise into single-pulse observations (e.g.~\citealt{Shannon2014}), we average the timing residuals (the difference between the observed pulse arrival times and those expected from a timing model) into $2 \, \se$ bins and center the data around the glitch epoch $t_g$ given by \citet{Palfreyman2018}. We subsequently determine the timing residuals corresponding to our predicted spin-up. Provided that residuals are small, they are proportional to $-\phi$. As demonstrated by Figure~\ref{fig:delta_nu} (right panels), the residuals thus start at zero and become increasingly negative with time, reproducing the characteristic glitch signature. The observation however reveals positive (approximately constant) timing residuals after $t_g$. In concordance with the magnetospheric changes accompanying the glitch \citep{Palfreyman2018}, we do not interpret this as a spin-down of the pulsar but instead as a phase shift. We include this by applying a constant shift $\Delta t$ to the residuals, so that theoretical predictions and observation agree at $t = 0$. A comparison between the resulting timing residuals as well as the cumulative residuals is shown in Figure~\ref{fig:data_comparison} for the first $120 \, \se$ after the glitch based on $\CB_{\rm core} \approx 5 \times 10^{-5}$.

\begin{figure}[b!]
\centering
\includegraphics[width = 0.47\textwidth]{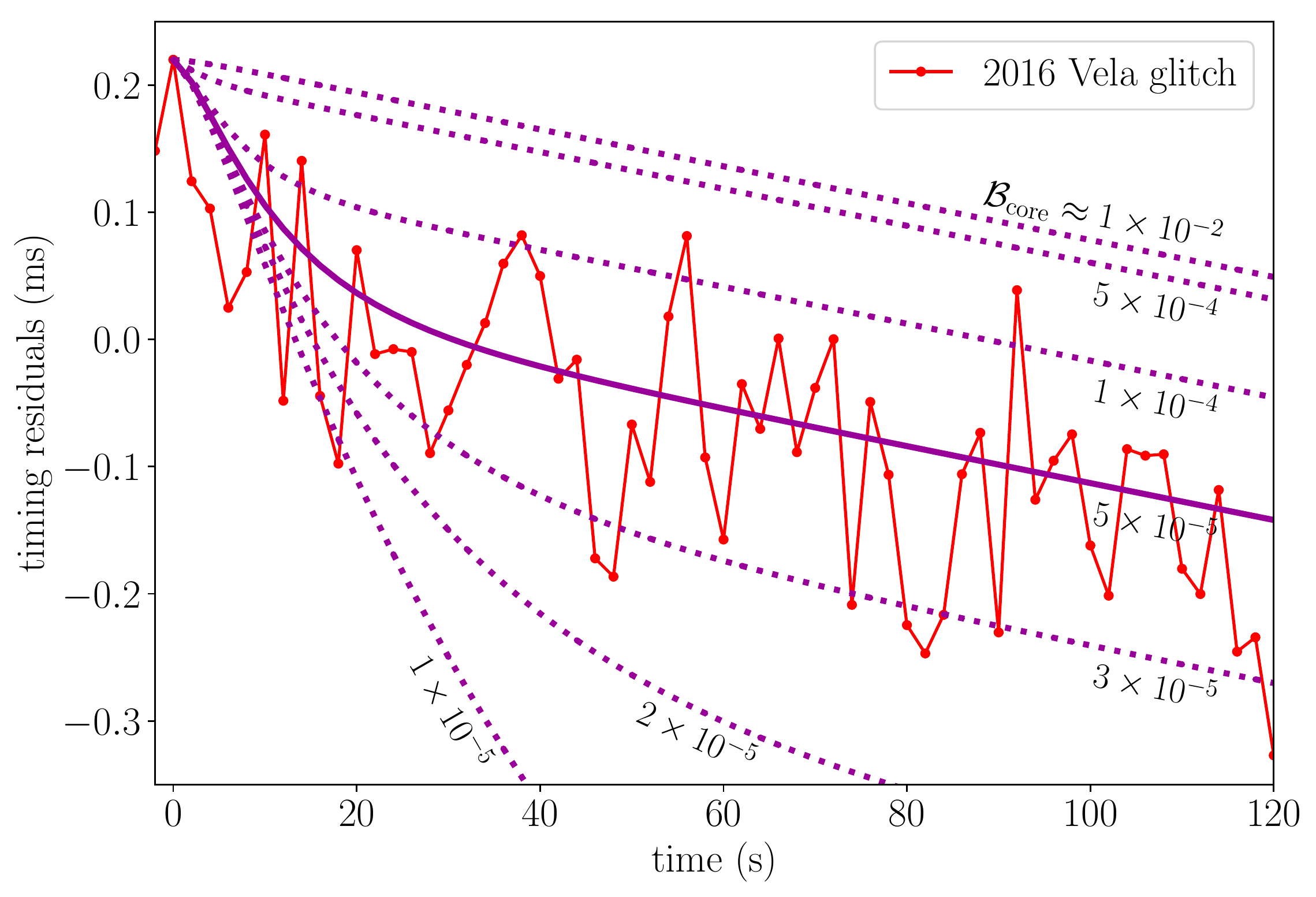}
\caption{\col{Comparison between the 2016 Vela glitch data averaged into $2 \, \se$ bins and theoretical predictions calculated for the crustal drag profile (A) and a varying crust-core mutual friction strength $\CB_{\rm core}$, as labeled in the figure.}}
\label{fig:data_comp_Bcore}
\end{figure}

Figure~\ref{fig:data_comparison} \col{highlights that the initial post-glitch response is rather insensitive to the crustal mutual friction profile provided that $\CB \gtrsim 10^{-3}$. We observe that models (A), (B) and the constant couplings $\CB \approx 10^{-1}, 10^{-2}, 10^{-3}$ fit the data similarly well. Only profile (C) (the weakest close to the crust-core interface) and $\CB \approx 10^{-4}$ decrease slower than what is observed, suggesting that strong mutual friction prevails in a sizable fraction of the inner crust. Together with the fiducial choice $\CB_{\rm core} \approx 5 \times 10^{-5}$ this ensures that a large portion of the superfluid's angular momentum can be transferred to the crust component before the core recoupling takes place. In order to illustrate the interplay of the crust and core coupling strengths in more detail, we focus on the strongest coupling profile (A) and determine the glitch response for a range of crust-core coupling coefficients. As depicted in Figure~\ref{fig:data_comp_Bcore}, the neutron star's rotational evolution is very sensitive to $\CB_{\rm core}$. Disagreement between the model predictions and data is amplified as soon as $\CB_{\rm core}$ diverges from the fiducial value: For stronger (weaker) mutual friction, the phase shifts become much smaller (larger), which results in smaller (larger) timing residuals. Our comparison thus suggests that the dominant core mutual friction mechanism covers a rather narrow range $3 \times 10^{-5} \lesssim \CB_{\rm core} \lesssim 10^{-4}$, as typical for electron scattering off magnetized vortices \citep{Alpar1984}.}
%


\section{Discussion}
\label{sec:discussion}

For the first time, we calculate mutual friction profiles resulting from Kelvin wave excitation for a realistic crust model and combine those with a simplified treatment of the crust-core coupling to develop a predictive model of the glitch rise. We find that density-dependent coupling affects the amount of angular momentum that can be exchanged on specific timescales and hence influences the glitch response of the crust. This illustrates that uncertainties in deriving the underlying $\CB$ and microscopic parameters have a crucial influence on observables.

We demonstrate that the $\CB$ profiles depend most sensitively on the assumed vortex-nucleus interaction. \col{Model (A) accounts for the contributions $E_{\rm s, l}$ included by \citet{Epstein1992}, which remain almost constant at high densities. This causes stronger drag and thus faster recoupling. For the profiles (B) and (C), we instead considered $E_{\rm p}$, which decreases significantly with density due to collective pinning, and results in longer coupling timescales. Nonetheless, differences remain between the glitch rise predictions based upon the formalisms of \citet{Epstein1992} and \citet{Jones1992} due to their respective assumptions on the interaction potentials and dissipation length scales (see Figure~\ref{fig:mf_crust})}.

Other microphysical parameters of the crust also play an important role. Whereas the composition itself does not vary significantly between different EoSs, our results are sensitive to superfluid parameters such as the energy gap and in principle entrainment, which we have neglected to keep our introductory analysis tractable. Strong entrainment would reduce the size of the crustal angular momentum reservoir, causing difficulties for the `crust-only' glitch framework \citep{Chamel2013, Andersson2012} (see however \citealt{Watanabe2017}). Future work will be needed to address how entrainment impacts on the initial glitch response. Our results are further strongly affected by the pinning strength. Calculations of these parameters rely on many assumptions and are very uncertain: Whereas \citet{Epstein1988} and \citet{Donati2006} employed a Ginzburg-Landau approach and semi-classical model, respectively, \citet{Avogadro2008} have examined the vortex-nucleus interaction using a quantum mean-field framework arriving at pinning energies of opposite signs. Future work is essential to reconcile these results. A correct description of vortex transport should also account for interactions with a nuclear pasta phase expected to be present close to the crust-core interface \citep{Ravenhall1983}. This high-density region carries the majority of the crustal mass and should strongly affect the post-glitch behavior. Real-time studies of the vortex-nucleus interaction \citep{Bulgac2013, Wlazlowski2016} could help to address this issue, but it remains unclear how this microscopic picture relates to the dynamics of a mesoscopic vortex communicating with many nuclei.

Finally, note that we based our model on the assumption that Kelvin wave excitations dominate the dissipation. Other processes, such as vortex coupling to lattice defects or impurities \citep{Harding1978}, could similarly alter the glitch response and their effects studied as outlined above once the mutual friction profile is known.

In addition to crustal microphysics, the shape of the glitch rise is crucially influenced by the relative strength between crust coupling and core mutual friction. The amount of angular momentum that the superfluid transfers to the crust before the core is recoupled controls the size of the phase shifts, providing the means to constrain the core physics. This plays an important role in comparing our predictive model with the first resolved glitch rise observation of the December 2016 Vela glitch \citep{Palfreyman2018}. Although a more detailed analysis will be needed to systematically study the impact of the underlying microscopic parameters on the glitch rise, our comparison points toward strong crustal mutual friction \col{satisfying $\CB \gtrsim 10^{-3}$} in combination with weaker core coupling \col{in the range $3 \times 10^{-5} \lesssim \CB_{\rm core} \lesssim 10^{-4}$. Such strengths as typical for electron scattering off the magnetized vortices \citep{Alpar1984, Andersson2006b}, but much weaker than the drag associated with excitations of vortex Kelvin waves in the neutron core \citep{Link2003, Haskell2014}. The absence of strong dissipation (characteristic for the regime where vortex-fluxtube interactions dominate the dynamics) could be explained if the protons do not form a type-II superconductor. Coupling dynamics in a type-I state are however rather uncertain \citep{Sedrakian2005, Jones2006b}. Furthermore, our predictive model only accounts for constant $\CB_{\rm core}$ values, and additional work incorporating density-dependent crust-core coupling would be needed to verify an absence of strong core friction.} Our conclusions were further based on the assumption that the Vela pulsar undergoes a shift in phase at the time of the glitch. Future observations will be required to confirm if this is justified and the phase shift is indeed a real feature of pulsar glitches. Upcoming facilities like the Square Kilometer Array will play an important role in this endeavor as they may allow the glitch rises of other sources to be observed \citep{Watts2014, Kramer2015}.


\acknowledgments

We thank the referee for suggesting an improved treatment of the microscopic vortex velocity and \citet{Palfreyman2018} for making their data publicly available. This work also benefited from discussions with Robert Archibald, Evan Keane and Toby Wood. V.~G.~is supported by a McGill Space Institute postdoctoral fellowship and the Trottier Chair in Astrophysics and Cosmology. A.~C.~is supported by an NSERC Discovery grant, is a member of the Centre de Recherche en Astrophysique du Qu\'ebec (CRAQ), and thanks Newcastle University for hospitality. N.~A.~acknowledges funding from STFC in the UK through grant number ST/M000931/1.

\vspace{5mm}

\software{IPython \citep{Perez2007}, Matplotlib \citep{Hunter2007}, NumPy \citep{Oliphant2006}, Pandas \citep{McKinney2010}, SciPy \citep{Jones2011}}


\end{document}